\documentclass[conference]{IEEEtran}
\IEEEoverridecommandlockouts
\usepackage{cite}
\usepackage{amsmath,amssymb,amsfonts}
\usepackage{algorithmic}
\usepackage{graphicx}
\usepackage{siunitx}
\usepackage{booktabs}
\usepackage{textcomp}
\usepackage{xcolor}
\usepackage[ruled]{algorithm2e}
\usepackage{bbding}
\usepackage{pifont}
\usepackage{fancyhdr}
\usepackage{soul}
\soulregister\ref7
\soulregister\cite7 
\soulregister\citep7 
\soulregister\citet7 

\usepackage{indentfirst}
\usepackage[ruled]{algorithm2e}
\def\BibTeX{{\rm B\kern-.05em{\sc i\kern-.025em b}\kern-.08em
    T\kern-.1667em\lower.7ex\hbox{E}\kern-.125emX}}

\begin{document}

\title{
Towards a High-performance and Secure Memory System and Architecture for Emerging Applications
}


\author{\IEEEauthorblockN{Zhendong Wang, 
Yang Hu (Advisor) {\{zhendong.wang, yang.hu4\}@utdallas.edu},
}
\IEEEauthorblockA{\textit{Electrical and Computer Engineering Department, The University of Texas at Dallas, Richardson, US}
}
}

\maketitle
\footnotetext[1]{The paper has been accepted by DAC'22 PhD PhD Forum.}
In the 5G era, diverse types of artificial intelligence (AI) and Internet of Things (IoT) applications emerge in our life, such as smart homes, virtual reality, and autonomous vehicles \cite{rtas22, dac22}. These applications typically impose diversified requirements in real deployments in terms of latency, privacy, security, etc., and stimulate the evolution and prosperity of heterogeneous computing. In this dissertation, heterogeneous computing indicates the scheme in which the different computing Processing Units (PUs) with differentiated computing capacities are effectively coordinated and managed to achieve computing gains. The representative PUs include CPU, Graphics Processing Unit (GPU) \cite{coop}, Field Programmable Gate Array (FPGA) \cite{iccd19}, Application-Specific Integrated Circuits (ASIC), and etc.

As GPU has become one of the most promising and prevalent platforms to deploy emerging AI-enabled applications, \textbf{this dissertation sets to discuss some key challenges and solutions of GPU-based heterogeneous system/architecture, especially the memory subsystem and management, matching the deployment requirements of emerging applications from the performance and security perspective.}

Regarding the challenges, the applications typically process huge volumes of data and computations and are memory-hungry, and can exhibit diverse computation properties and memory access patterns. In contrast, the GPU-based heterogeneous system/architecture, especially the GPU device, has limited memory capacity. Also, the CPU PU and GPU PU in the heterogeneous system have fundamentally different computing architectures and differentiated memory subsystems. Thus, there exists a "memory wall" caused by the mismatch between the diversified applications properties and the GPU-based system heterogeneity, which damages the applications' performance. On the other hand, applications face a variety of security and privacy risks during deployments. However, the GPU-based heterogeneous system, especially the memory subsystem, can expose multiple security vulnerabilities, damaging applications' privacy.

\begin{figure}[t]
\centering
            \includegraphics[width=\linewidth]{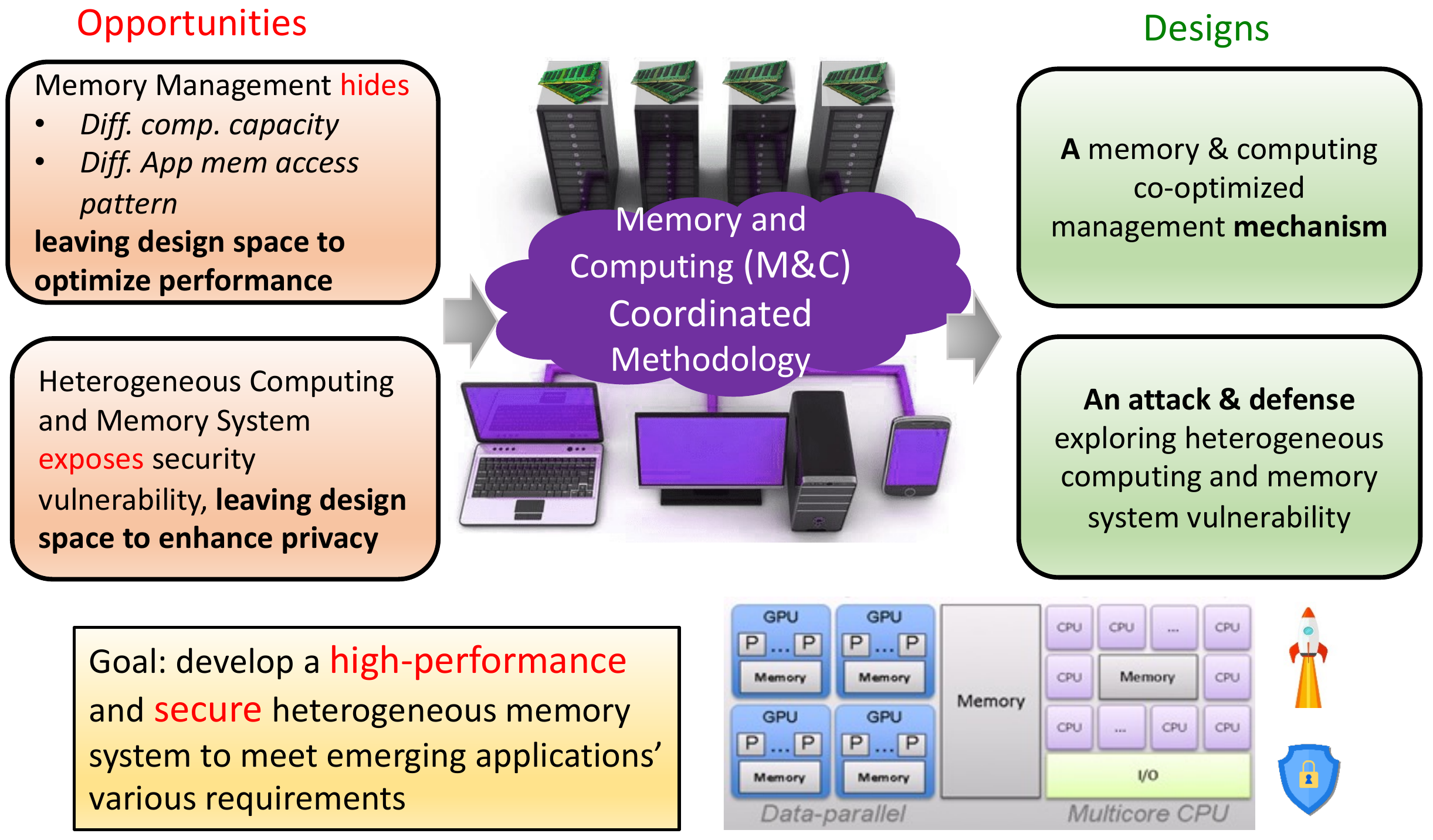}
            \vspace{-6mm}
            \caption{Overview of the dissertation: opportunities/challenges, proposed methodology and specific designs.}\label{design}
            \vspace{-8mm}
\end{figure}

\textbf{Due to the challenges of "memory wall" and security breach, the deployment of emerging AI-enabled applications in existing GPU-based heterogeneous systems does not thoroughly exploit the power of the system and does not match the applications' properties and requirements to the strengths of CPU/GPU architectures, which thus damages applications' performance and privacy.}

1) "Memory wall" damages application performance. 
In this dissertation, the "memory wall" indicates that the memory management (MM) mechanism of exiting GPU-based heterogeneous system is not aware of the mismatch between the diversified applications computation and memory access properties and the differentiated CPU/GPU computing and memory architectures/characteristics, 
thus damaging applications' performance (e.g., response time).


On one hand, existing memory management is not aware of the differentiated computing and memory capacities between CPU and GPU and does not fully explore the CPU/GPU differentiated architecture and strengths to optimize applications' performance. 
On the other hand, the existing GPU-based heterogeneous systems can provide multiple memory management models (e.g., the representative copy-then-execute model and unified memory model. 
We observe that the diverse applications can cause different latency and memory footprints under the different memory management models due to the different computation properties and memory access patterns of the applications. 
However, existing memory management in the GPU-based heterogeneous systems simply adopts the copy-then-execute model for all applications and is not aware of the characteristics of the available different memory management models (i.e., their advantages and disadvantages) as well as different applications’ properties.
In other words, the existing memory management does not thoroughly exploit the characteristics and power of the GPU heterogeneous system/architecture to optimize applications' performance (e.g., adapt the different memory management models to different applications to reduce applications' latency, memory footprint, etc.).

2) Security breach damages application privacy. 
The rising popularity of AI and machine learning (ML) technologies, especially DNNs, benefits various application domains, such as computer vision, speech recognition, etc.
These DNNs models are deemed confidential due to their unique value in the expensive training efforts, privacy-sensitive training data, and proprietary network characteristics.  
Consequently, the model value raises the incentive for an adversary to steal the model for profits, such as the representative model extraction attack. 
The model extraction attack not only destroys the confidentiality of a model but also benefits further adversarial attacks. 

Nowadays, GPU has become the dominant hardware to deploy DNN applications. 
Also, the considerable memory footprint of DNN-based workloads and ever-increasing requirements of programming flexibility has pushed the GPU memory management on the verge of a major shift from the traditional copy-then-execute (CoE) model to the unified memory (UM) model. 
Prior studies show that the DNN execution can exhibit some runtime behaviors (e.g., memory traffic) and cause hardware/architecture-level information leakage (e.g., memory transactions/throughput), and the extraction attack can benefit from exploring the hardware/architecture-level information leakage to improve the attack performance (e.g., extraction accuracy). 
Essentially, the hardware/architecture-level information leakage provides an attack surface for the adversary.
However, the commonly collected hardware/architecture-level information in GPU systems mainly targets the copy-then-execute memory model and can become less effective when it is still used in the extraction attacks under the populating unified memory model. Therefore, to achieve an effective attack, the adversary pursuits for new hardware/architecture-level information leakage under unified memory, which can expose new attack surfaces of the GPU unified memory system and damages applications' privacy.



Facing the challenges of deploying emerging AI-enabled applications in the GPU-based heterogeneous system, we obverse such opportunities, as shown in Fig. \ref{design}. 1) As existing memory management mechanism hides the differentiated computing and memory characteristics of the heterogeneous system and the diversified applications' computation and memory access properties, damaging applications' performance, this leaves us an opportunity to optimize applications' performance by effectively coordinating the computing and memory capacities and characteristics of the system. 2) As GPU-based heterogeneous system, especially the memory subsystem and management, exposes security vulnerabilities, damaging applications' privacy, it leaves us an opportunity to further explore the potential memory vulnerabilities and enhance applications' privacy by effectively reinforcing the memory security and performance. 

Motivated by the opportunities, we \textbf{propose a memory and computing coordinated methodology to thoroughly exploit the characteristics and capabilities of the GPU-based heterogeneous system to optimize applications' performance and enhance applications' privacy}.
Specifically, we have four designs. 1) We propose a task-aware and dynamic memory management mechanism to co-optimize applications' latency and memory footprints, especially in multitasking scenarios \cite{coop}. 2) We propose a novel latency-aware memory management framework that analyzes the application characteristics and hardware features to reduce applications' initialization latency and response time \cite{enable20, hotedge20}. 3) We develop a new model extraction attack that explores the vulnerability of the GPU unified memory system to accurately steal private DNN models. 4) We propose a CPU/GPU Co-Encryption mechanism that can defend against a timing-correlation attack in an integrated CPU/GPU platform to provide a secure execution environment for edge applications \cite{ats21}.

We seek to develop a high-performance and secure memory system and architecture in GPU heterogeneous platforms to deploy emerging AI-enabled applications efficiently and safely. 

\noindent{\large\bfseries Supporting Paper:\par} Soroush Bateni*, \textbf{Zhendong Wang*}, Yuankun Zhu, Yang Hu and Cong Liu. “Co-Optimizing Performance and Memory Footprint Via Integrated CPU/GPU Memory Management, an Implementation on Autonomous Driving Platform” (* equal contribution). 
IEEE Real-Time and Embedded Technology and Applications Symposium (RTAS, CSRanking Top), 2020.\par

\textbf{Description}:Cutting-edge embedded system applications, such as self-driving cars and unmanned drone software, are reliant on integrated CPU/GPU platforms for their DNNs-driven workload, such as perception and other highly parallel components. In this work, we set out to explore the hidden performance implication of GPU memory management methods of integrated CPU/GPU architecture. Through a series of experiments on micro-benchmarks and real-world workloads, we find that the performance under different memory management methods may vary according to application characteristics. Based on this observation, we develop a performance model that can predict system overhead for each memory management method based on application characteristics. Guided by the performance model, we further propose a runtime scheduler. By conducting per-task memory management policy switching and kernel overlapping, the scheduler can significantly relieve the system memory pressure and reduce the multitasking co-run response time. We have implemented and extensively evaluated our system prototype on the NVIDIA Jetson TX2, Drive PX2, and Xavier AGX platforms, using both Rodinia benchmark suite and two real-world case studies of drone software and autonomous driving software.



\begin{thebibliography}{00}
\bibitem{coop} Bateni, S., Wang, Z., Zhu, Y., Hu, Y., \& Liu, C. (2020, April). Co-optimizing performance and memory footprint via integrated cpu/gpu memory management, an implementation on autonomous driving platform. In 2020 IEEE Real-Time and Embedded Technology and Applications Symposium (RTAS) (pp. 310-323). IEEE.

\bibitem{enable20} Wang, Z., Jiang, Z., Wang, Z., Tang, X., Liu, C., Yin, S., \& Hu, Y. (2020). Enabling Latency-Aware Data Initialization for Integrated CPU/GPU Heterogeneous Platform. IEEE Transactions on Computer-Aided Design of Integrated Circuits and Systems, 39(11), 3433-3444.

\bibitem{hotedge20} Wang, Z., Wang, Z., Liu, C., \& Hu, Y. (2020). Understanding and tackling the hidden memory latency for edge-based heterogeneous platform. In 3rd USENIX Workshop on Hot Topics in Edge Computing (HotEdge 20).

\bibitem{ats21} Wang, Z., Wang, R., Jiang, Z., Tang, X., Yin, S., \& Hu, Y. (2021, November). Towards a Secure Integrated Heterogeneous Platform via Cooperative CPU/GPU Encryption. In 2021 IEEE 30th Asian Test Symposium (ATS) (pp. 115-120). IEEE.

\bibitem{rtas22} Liu, S., Wang, J., Wang, Z., Yu, B., Hu, W., Liu, Y., ... \& Hu, Y. (2022, May). Brief industry paper: The necessity of adaptive data fusion in infrastructure-augmented autonomous driving system. In 2022 IEEE 28th Real-Time and Embedded Technology and Applications Symposium (RTAS) (pp. 293-296). IEEE.

\bibitem{dac22} Zeng, X., Wang, Z., \& Hu, Y. (2022). Enabling Efficient Deep Convolutional Neural Network-based Sensor Fusion for Autonomous Driving. arXiv preprint arXiv:2202.11231.

\bibitem{iccd19} Maor, G., Zeng, X., Wang, Z., \& Hu, Y. (2019, November). An FPGA implementation of stochastic computing-based LSTM. In 2019 IEEE 37th International Conference on Computer Design (ICCD) (pp. 38-46). IEEE.






\end{thebibliography}
\end{document}